\documentclass[lettersize,journal]{IEEEtran}
\usepackage{graphicx}
\usepackage{adjustbox}
\usepackage{cite}
\usepackage{array}
\usepackage{dblfloatfix}
\usepackage{float}
\usepackage{amssymb}
\usepackage{multirow}
\usepackage{tikz}
\usepackage{tabularx}
\usepackage{booktabs}
\usepackage{adjustbox}
\usepackage{balance}
\usepackage{placeins}
\usepackage{afterpage} 
\usepackage{makecell}
\usepackage[symbol]{footmisc} 
\usepackage{enumitem}

\usepackage{algorithm}
\usepackage{algorithmic}

\usepackage{xcolor}
\usepackage{soul}
\sethlcolor{yellow}

\usepackage{cuted} 
\usepackage{url}

\ifCLASSINFOpdf
\else
\fi

\usepackage{amsfonts}
\usepackage{amsmath} 
\usepackage[caption=false,font=normalsize,labelfont=sf,textfont=sf]{subfig}
\usepackage{textcomp}
\usepackage{url}
\usepackage{verbatim}
\def\BibTeX{{\rm B\kern-.05em{\sc i\kern-.025em b}\kern-.08em
    T\kern-.1667em\lower.7ex\hbox{E}\kern-.125emX}}

\begin{document}

\title{D-Legion: A Scalable Many-Core Architecture for Accelerating Matrix Multiplication in Quantized LLMs}

\author{Ahmed~J.~Abdelmaksoud, 
        Cristian Sestito, \textit{Member, IEEE},
        Shiwei Wang, \textit{Senior Member, IEEE},
        and Themis~Prodromakis, \textit{Senior Member, IEEE}
\thanks{This work was supported by EPSRC FORTE Programme (Grant No.
EP/R024642/2), the RAEng Chair in Emerging Technologies (Grant
No. CiET1819/2/93), and EPSRC AI Hub for Productive Research and Innovation in eLectronics (APRIL) (Grant No. EP/Y029763/1)}

\thanks{Ahmed J.Abdelmaksoud, Cristian Sestito, Shiwei Wang and Themis Prodromakis are with the Centre for Electronics Frontiers, Institute for Integrated Micro and Nano Systems, School of Engineering, The University of Edinburgh, EH9 3BF, Edinburgh, United Kingdom
(e-mails: a.j.abdelmaksoud@ed.ac.uk; csestito@ed.ac.uk; shiwei.wang@ed.ac.uk; t.prodromakis@ed.ac.uk).}}%


\maketitle

\begin{abstract}
The performance gains obtained by large language models (LLMs) are closely linked to their substantial computational and memory requirements. 
Quantized LLMs offer significant advantages with extremely quantized models, motivating the development of specialized architectures to accelerate their workloads.
This paper proposes D-Legion, a novel scalable many-core architecture, designed using many adaptive-precision systolic array cores, to accelerate matrix multiplication in quantized LLMs.
The proposed architecture consists of $L$ Legions where each Legion has $C$ systolic arrays. 
D-Legion supports multiple computation modes, including quantized sparse and dense matrix multiplications.
The block structured sparsity is exploited within a fully-sparse, or partially-sparse windows. 
In addition, memory accesses of partial summations (psums) are spatially reduced through parallel accumulators. 
Furthermore, data reuse is maximized through optimized scheduling techniques by multicasting matrix tiles across the Legions. 
A comprehensive design space exploration is performed in terms of Legion/core granularity to determine the optimal Legion configuration.
Moreover, D-Legion is evaluated on attention workloads from two BitNet models, delivering up to 8.2× lower latency, up to 3.8× higher memory savings, and up to 3× higher psum memory savings compared to state-of-the-art work. 
D-Legion, with eight Legions and 64 total cores, achieves a peak throughput of 135.68 TOPS at a frequency of 1 GHz.
A scaled version of D-Legion, with 32 Legions, is compared to Google TPUv4i, achieving up to 2.5× lower total latency, up to 2.3× higher total throughput, and up to 2.7× higher total memory savings.

\end{abstract}


\IEEEpeerreviewmaketitle

\section{Introduction}

\IEEEPARstart{T}{he} rapid rise of LLMs, based on the foundational Transformer architecture, marks a turning point in achieving state-of-the-art performance in natural language processing (NLP) tasks \cite{naveed2025comprehllms, changllmsurvey}.
The attention mechanism is the core innovation that makes LLMs capable of sophisticated language understanding by capturing diverse representational subspaces; however, attention layers are heavily dependent on large-scale matrix multiplications.
As a result, the obtained performance gains entail substantial memory and computational resources. Unlike the one-time expense of training, inference costs accumulate with growing demands. This inherent trade-off between model performance and computational efficiency has catalyzed ongoing research into model compression techniques, aimed at achieving comparable predictive accuracy while minimizing resource requirements for inference operations.

Model compression involves a range of methods including pruning and quantization to reduce model size, while preserving performance, thereby lowering memory footprint and energy consumption \cite{tang2024survey}.
Pruning removes redundant or less significant parameters, resulting in smaller, faster, and more efficient models.
Quantization further enhances efficiency by representing model weights and activations with lower precision.
Post-training quantization (PTQ) and quantization-aware training (QAT) are two complementary quantization strategies \cite{jacob2018quantization}. 
PTQ applies quantization to a pre-trained model without altering the weights during training. While PTQ does not require changes to the training phase, it may lead to accuracy degradation since the model is not optimized for the reduced precision during its learning phase. 
On the other hand, QAT integrates quantization into the training phase, allowing the model to compensate for the reduced the precision, especially for heavily quantization schemes. This results in minimal loss of accuracy compared to PTQ.

BitNet models push the boundaries of model efficiency by employing QAT with extreme quantization approaches that explore 1-bit and ternary weights, demonstrating that models with ultra-low-precision can maintain competitive performance while drastically reducing computational and memory requirements \cite{diao2023bitnet}. BitNet defines a new methodology for training new generations of LLMs that are cost-effective and opens a new research area for designing specialized hardware architectures optimized for 1-bit LLMs.

Graphics processing units (GPUs) have served as the primary hardware platform for AI development due to their high degree of parallelism and strong performance. However, as model sizes and inference demands have grown, GPUs have struggled to provide the scalability and efficiency required to sustain the growing size of AI workloads, especially Transformers and LLMs. 
This limitation has motivated the development of specialized hardware accelerators designed specifically to handle different workload patterns \cite{kim2023full}. 
Following the GPUs, Google introduced multiple generations of tensor processing units (TPUs) that deliver higher efficiency for training and deploying large-scale workloads \cite{jouppi2021tpuv4i}. Systolic arrays (SAs) are the core compute architecture behind TPUs, a spatial architecture well-suited for compute-intensive matrix multiplications and capable of alleviating memory bandwidth bottlenecks by maximizing data reuse. As LLM and Transformer compression techniques continue to evolve, they introduce new opportunities and requirements for hardware efficiency, particularly during inference. This progression underscores the need for even more specialized hardware architectures that can fully exploit the computational and memory advantages of aggressively quantized LLMs, enabling high-throughput, energy-efficient deployment at scale.

A wide range of specialized hardware accelerators has emerged to efficiently support large-scale matrix multiplications central to attention workloads.
Moreover, the evolution of quantization methods paves the way to develop adaptive-precision architectures that dynamically adjust precision and explores the sparsity opportunities for energy-efficient processing \cite{ibrahim2022taxonomy, liu2025multi, yang2022dtatrans, xu2021hesa, yang2024trapezoid, he2020sparsetpu}. 
Complementary to these approaches, multi-core architectures have been proposed to provide scalable and high-performance solutions \cite{abts2020think, genc2021gemmini, samajdar2022sara, tang2024msa2, liu2022dynamic, chen2024efficient, choi2023enabling, ghodrati2020planaria, yuzuguler2023scale}.
However, a gap remains in unifying these directions into a single architecture. 
Addressing this gap requires developing a new hardware architecture that combines scalable many-core design with adaptive-precision and sparsity-aware execution, tailored for accelerating attention workloads in quantized Transformers and LLMs. 

In this paper, we present D-Legion, a novel scalable many-core architecture, designed to accelerate matrix multiplication in quantized LLMs. The main contributions of this work are highlighted as follows:
\begin{itemize}
  \item The proposed architecture consists of many systolic array cores, grouped in Legions, with adaptive-precision processing elements (PEs), providing up to 4× higher computational density for quantized attention workloads.
  \item D-Legion spatially reduces memory accesses of psums using parallel accumulators per each Legion.
  \item Data reuse is maximized through optimized scheduling techniques and flexible network-on-chip design by multicasting the matrix tiles during the processing of attention layers.
  \item A comprehensive hardware design space exploration is performed, showing Legion/core granularity analysis to determine the optimal architecture configuration along with analytical modeling.
  \item D-Legion is evaluated and compared to state-of-the-art work using attention workloads from two BitNet models.
  \item  Scalability analysis is conducted for the proposed architecture, showing that the number of Legions can be increased to up to 64 Legions, delivering up to 1085.44 TOPS. In addition, D-Legion, with 32 Legions, is compared to Google TPUv4i using attention workloads from two BitNet models.
  
\end{itemize}

This paper is organized as follows: Section II discusses the related work and background. Section III presents the design space exploration. Section IV presents D-Legion architecture. Section V shows the evaluation and results. Finally, section VI concludes the work.

\section{Related Work \& Background}

\subsection{WS, DiP, and ADiP Architectures}

SAs are spatial architectures consisting of a 2D grid of interconnected PEs, and designed to improve the computational efficiency and data reuse \cite{kung1982systolic, xusasurvey}. The input weights and activations flow among the PEs in a wave-fashion manner, while the communication with the array occurs only at the boundary PEs. This approach improves local data reuse by increasing the intensity of operations per memory fetch.
Each PE typically consists of a multiplication and accumulation (MAC) unit along with additional registers.
There are well-known SAs for matrix multiplication acceleration, such as weight-stationary (WS), output-stationary (OS), and input-stationary (IS). 
However, WS systolic array is widely-used as it keeps the weights stationary for the longest processing time. 

WS architecture consists of a grid of $N$×$N$ PEs and two first-in-first-out (FIFO) groups for input and output synchronization, adopted by many accelerators, such as Google TPUs \cite{jouppi2021tpuv4i}. WS SAs offer higher data reuse by keeping the weight matrix stationary, and moving the input matrix and psums among the PEs.  
However, WS instantiates input and output FIFOs to synchronize the data. 
The input FIFOs map the input matrix to the PEs in a diagonal wave fashion, while the output FIFOs synchronize partial summations. 
The synchronization FIFOs significantly add overheads in terms of area, power, energy consumption, and latency during matrix multiplication. In addition, overall PE utilization is reduced as the computations propagate as a diagonal wave, degrading the performance.

DiP is a systolic array architecture designed to eliminate the input and output synchronization FIFOs required by conventional WS architecture \cite{abdelmaksoud2025dip}. 
Each PE consists of a conventional MAC unit with an INT8 multiplier and register units.
DiP relies on two key improvements: the diagonal movement of the input matrix, and the weight matrix permutation. In the DiP dataflow, weight permutation is statically performed by shifting and rotating each column by its column index.
Eliminating input/output synchronization FIFOs offers up to 2× better energy efficiency per area, and improves latency, throughput, and PE utilization. 

Furthermore, ADiP is adaptive-precision systolic array architecture designed to efficiently process quantized matrix multiplication workloads. 
ADiP consists of $D$×$D$ reconfigurable PEs, in addition to shared shifters and accumulators. 
ADiP offers up to 4× higher computational density with 4× higher memory efficiency compared to DiP architecture for quantized workloads \cite{abdelmaksoud2025adip}.
ADiP inherits DiP dataflow with diagonal input movement and weight permutations.
Each reconfigurable PE integrates a set of 16 2-bit multipliers, arranged into four groups with their internal accumulators. The configuration of the reconfigurable PE is selected to process 8-bit×8-bit (8b×8b) multiplications with a latency of one cycle. Correspondingly, the computational throughput for 8-bit×2-bit (8b×2b) multiplications is quadrupled. 

\subsection{Background on Attention Layers}

\begin{figure}[b]
  \centering
  \includegraphics[width=\linewidth, keepaspectratio]{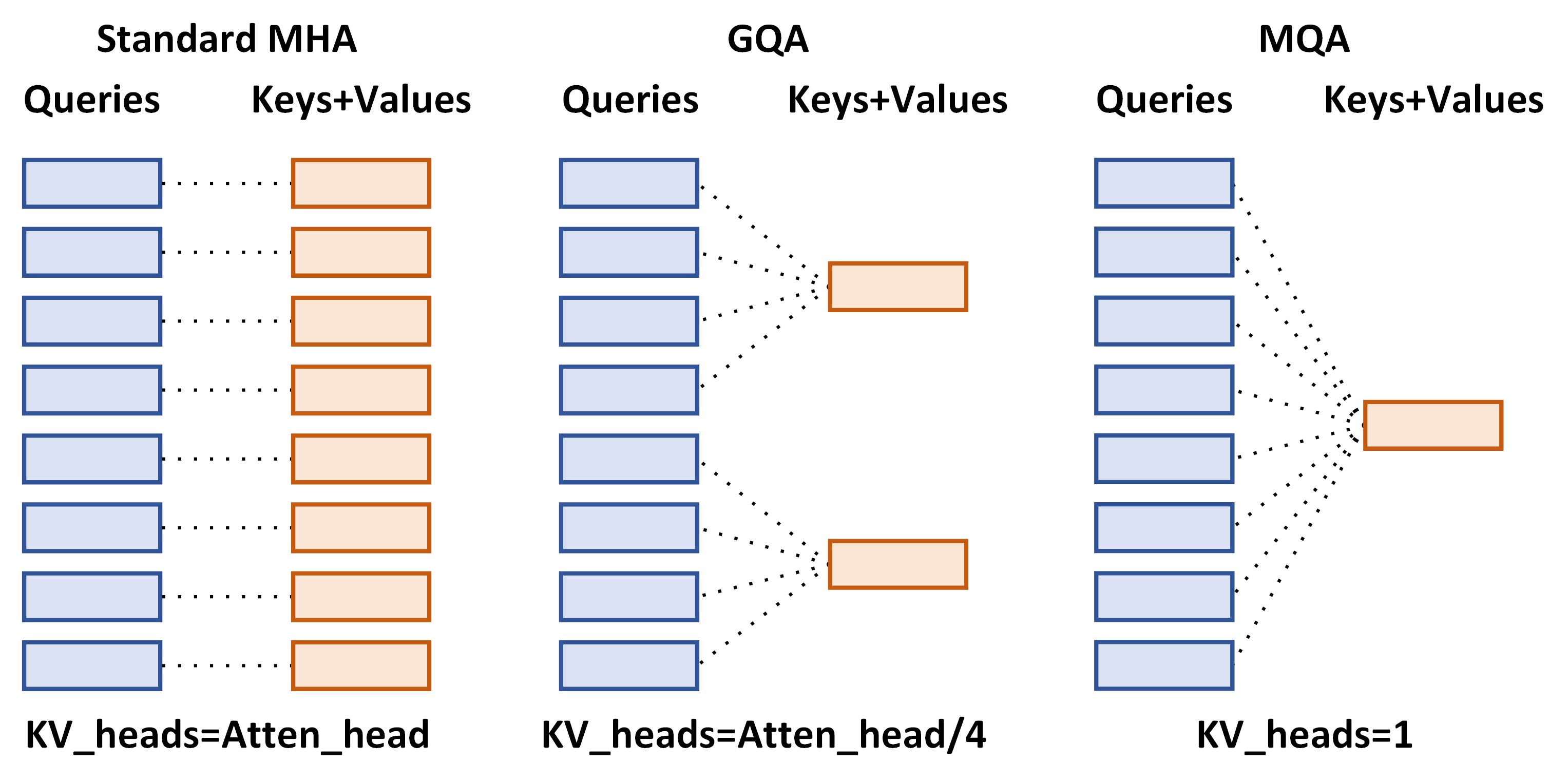} 
  \caption{Different attention layer types, including standard multi-head attention (MHA), grouped query attention (GQA), and multi-query attention (MQA).}
  \label{Fig.1}
\end{figure}

Multi-head attention (MHA) is the central computational primitive of modern Transformer models. MHA captures diverse contextual dependencies through parallel attention heads \cite{vaswani2017attention}.
Because MHA stores $H$ independent key-value (KV) matrices, its cache usage scales with both head count and sequence length, quickly becoming a hardware bottleneck during inference.
The adoption of grouped-query attention (GQA) and multi-query attention (MQA) arises from the escalating memory and bandwidth demands of maintaining per-head KV caches in standard MHA \cite{ainslie2023gqa}. GQA mitigates this by grouping heads to share only $G$ KV projections, significantly shrinking cache size and memory accesses with minimal loss of representational diversity, as shown in Fig. \ref{Fig.1}. 
MQA extends grouping and factorization across Q, K, and V. It effectively pushes the principle of KV sharing toward the extreme case, producing a single shared KV representation for all heads. 
However, increasing the degree of KV sharing typically introduces measurable accuracy loss, since reducing $H$ KV streams to $G$ or even one narrows the model’s representational space. 
GQA has been adopted in several large-scale models, such as BitNet model \cite{diao2023bitnet}. 

The practical importance of GQA and MQA emerges strongly at the hardware-level. As fewer KV matrices are stored, off-chip memory accesses are reduced, which alleviates the main bottleneck in modern accelerators. KV compression improves cache locality and permits attention layers to retain a larger fraction of their working set in high-speed on-chip memory. Furthermore, grouped attention mechanisms provide substantial reductions in inference latency and energy consumption.

\section{Design Space Exploration}

This section presents the design space exploration and analytical modeling to select the optimal core configuration for each Legion in terms of input bandwidths, time to full utilization (TFU), latency, and psum memory access for attention workloads from BitNet LLM model.

The architectural philosophy behind D-Legion involves decomposing a large systolic array core into smaller cores with additional accumulators to spatially reduce psums. The cores are grouped in a legion-centric fashion, wherein each Legion acts as a standalone accelerator to accelerate a unique matrix multiplication workload. This modular organization provides higher computational throughput, memory efficiency, and energy savings. 
The proposed architecture consists of $L$ Legions, where each Legion has $C$ adaptive-precision systolic array (ADiP) cores, each with \textit{D×D} reconfigurable PEs. 

Each ADiP core deploys reconfigurable PEs to efficiently accelerate quantized matrix multiplication workloads \cite{abdelmaksoud2025adip}.
Matrix tiles ($MT$, $KT$, $NT$) are calculated based on \eqref{eq:1} in terms of workload dimensions of ($M$, $K$, $N$). Correspondingly, end-to-end matrix multiplication latency for each Legion of $C$ ADiP cores is derived based on \eqref{eq:2}. 
D-Legion latency is the same as Legion latency because all Legions operate simultaneously, delivering linear-scale throughput with the same Legion latency.
Furthermore, TFU is a metric that measures the latency in cycles required to reach full utilization of PEs, as shown in \eqref{eq:3}. Additionally, this metric captures the overhead for weight tiles loading every iteration.

{\small
\begin{equation}
MT = \left\lceil \frac{M}{D} \right\rceil,\qquad
KT = \left\lceil \frac{K}{C \times D} \right\rceil,\qquad
NT = \left\lceil \frac{N}{R \times D} \right\rceil
\tag{1}\label{eq:1}
\end{equation}
}

\begin{equation}
  \mathit{Latency_{Legion}} = KT \times NT \times (D \times (MT + 1) + P) + D
\tag{2}\label{eq:2}
\end{equation}

\begin{equation}
    \mathit{TFU_{ADiP}} = D \tag{3}\label{eq:3}
\end{equation}

\noindent where the notations:
\begin{itemize}
    \item $D$: Systolic array row/column size (number of PEs)
    \item $M, K, N$: Matrix dimensions
    \item $MT, KT, NT$: Matrix tiles
    \item $R$: Acceleration ratio for quantized workloads, equals 1 and 4 for 8b×8b and 8b×2b operations, respectively.
    \item $P$: Number of pipeline stages
\end{itemize}

\begin{figure*}[t] 
  \centering
  \includegraphics[width=\textwidth, keepaspectratio]{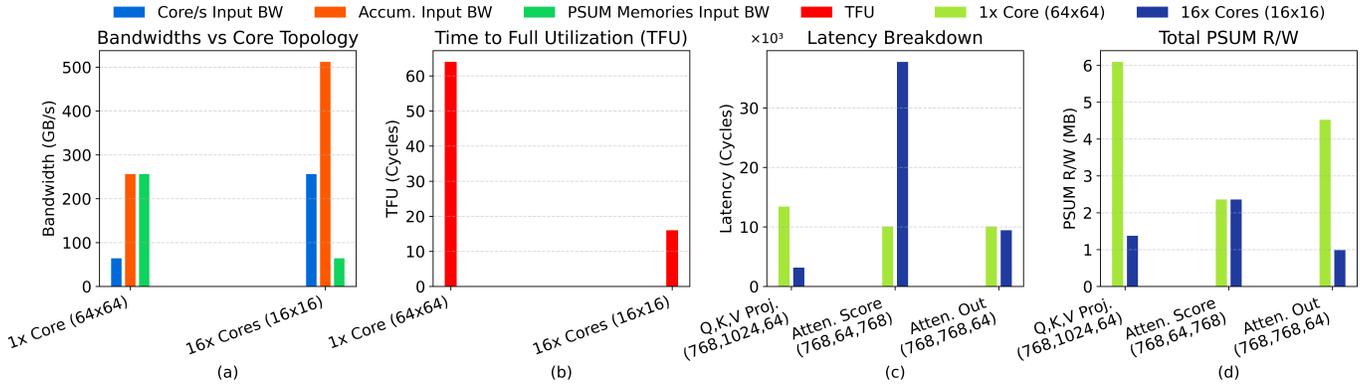} 
  \caption{A comprehensive analysis of single large core versus many smaller cores with the same number of PEs. (a) Input bandwidths of core(s), accumulation, and psum memories versus core topology. (b) TFU for each configuration. (c) Total Number of PEs per each configuration. (d) Latency breakdown across the analyzed configurations based on attention workloads, respectively.}
  \label{Fig.2}
\end{figure*}

\begin{figure*}[b] 
  \centering
  \includegraphics[width=\textwidth, keepaspectratio]{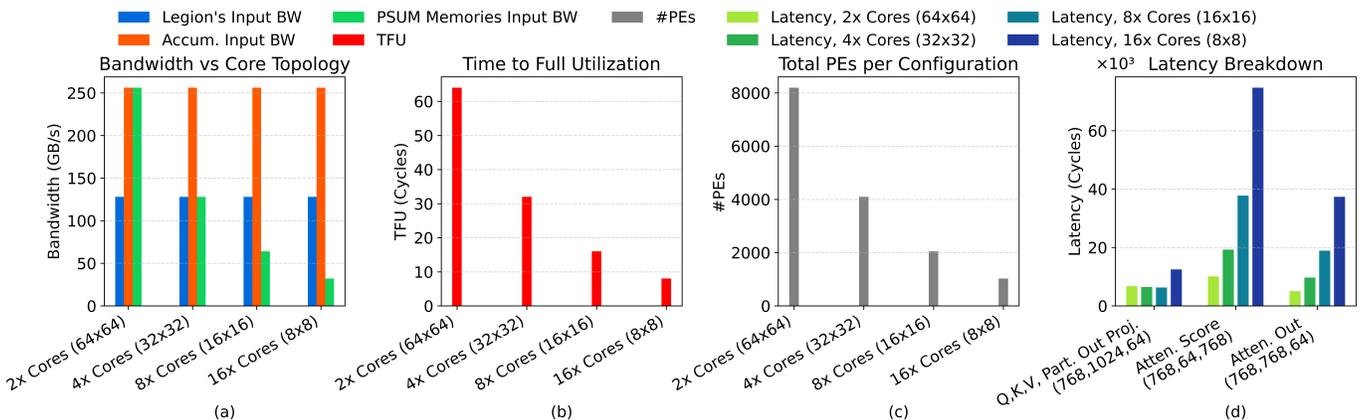} 
  \caption{Granularity analysis of the cores per Legion. (a) Input bandwidths of the Legion, accumulators, and psum memories versus core topology. (b) TFU per each Legion's configuration. (c) Total PEs per each Legion's configuration. (d) Latency breakdown across different Legion configurations based on attention workloads.}
  \label{Fig.3}
\end{figure*}

\subsection{Analysis of One Large Core vs Many Smaller Cores}

A comprehensive analysis of a single large core 64×64 versus multiple smaller cores 16×16×16 with the same total number of PEs is shown in Fig. \ref{Fig.2}. The large core is split into smaller cores that spatially reduce psum memory access and bandwidth requirements. 
First, Fig. \ref{Fig.2}(a) shows the input bandwidths of the core(s), accumulators, and psum memories versus core topology. The spatial-core configuration requires 4× higher input bandwidth, but 4× lower memory bandwidth than the single-core configuration. 
Second, TFU for the spatial-core configuration is 4× lower than the single-core configuration, as shown in Fig. \ref{Fig.2}(b). 
The latency and memory access breakdowns are studied based on workloads form BitNet LLM model \cite{wang20241}, as shown in Fig. \ref{Fig.2}(c) and (d). 
The head size of BitNet model are selected with 64, as a corner case of workloads processing. 
The latency of QKV projection workloads (with 2-bit weights) is reduced by 4× for the spatial-core configuration compared to the single-core configuration, as ADiP cores accelerate these workloads 4× along matrix dimension $N$. 
However, the latency of the attention score workload for the single-core configuration is 4× lower than the spatial-core configuration, as 75\% of the cores become unutilized. 
In addition, the latency of the attention output workload is similar, as both configurations perform the computations in dense mode with INT8 operations with the same number of PEs.  
Moreover, the spatial-core configuration achieves lower memory access for psums, as the cores spatially reduce psums before saving it to psum memory, as shown in Fig. \ref{Fig.2}(d). The first and third workloads achieve 4× lower psum memory access, while the second workload achieves similar memory access.

\subsection{Cores Granularity Analysis}

A comprehensive analysis of the granularity of cores per Legion is studied to select the optimal configuration in terms of the core size and the number of cores per Legion, as shown in Fig. \ref{Fig.3}. 
The analysis includes four different Legion configurations: 2×64×64, 4×32×32, 8×16×16, and 16×8×8.
The cores spatially reduce psums using parallel accumulators.
First, Fig. \ref{Fig.3}(a) shows the input bandwidths of the Legion, accumulators, and psum memories versus core topology. The configurations use the same Legion input bandwidth; however, psum memory bandwidth is gradually reduced as core size decreases.
Second, Fig. \ref{Fig.3}(b) shows that TFU is related to the core size in each configuration. Configurations with smaller core have better TFU than those with larger cores. 
While the previous analysis used the same number of PEs for both configurations, this analysis uses the same input bandwidth per Legion; however, the number of PEs decreases when using configurations with more cores and smaller core sizes, as shown in Fig. \ref{Fig.3}(c). 
In addition, Fig. \ref{Fig.3}(d) shows the latency breakdown across different Legion configurations based on the same workloads used in the first analysis. 
The latency for QKV projection workloads is similar for the first three configurations, as these workloads are bounded by the head size of 64. As ADiP cores accelerate the processing by 4× along matrix dimension $N$, the core configuration of 16×16 is optimal for these workloads. For the second and third workloads, the latency increase is related to the reduction in the number of PEs per configuration. 
Memory access is the same for all configurations.

\begin{figure}[b]
  \centering
  \includegraphics[width=\linewidth, keepaspectratio]{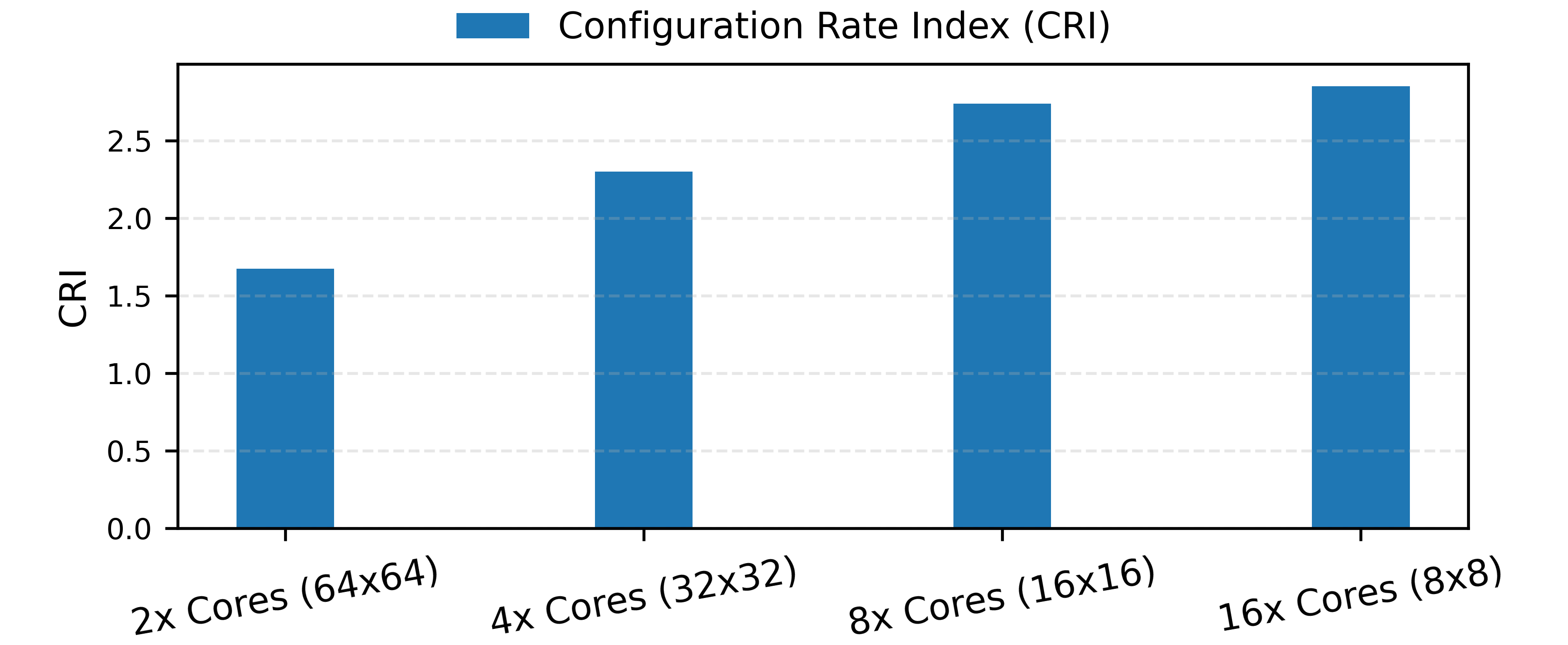} 
  \caption{Per-Legion configuration rate index (CRI), evaluating each Legion configuration to select the optimal number of cores and core size.}
  \label{Fig.4}
\end{figure}

Finally, configuration rate index (CRI) is introduced to select the optimal Legion configuration in terms of number of cores and core size. 
CRI considers the input bandwidths of the Legion, TFU, and the mean matrix multiplication latency for the selected corner case workloads. The breakdown of CRI across different configurations is shown in Fig. \ref{Fig.4}. 
The projection workloads with a head size of 64 are accelerated 4× with a configuration with core size of 16×16; however, the configuration with a core size of 32×32 loses the benefit of processing 4× higher number of workloads, and is limited to only 2× speedup. As a result, the configurations with 16×16 and 8×8 cores are more optimal than the ones with 64×64 and 32×32 cores.
Correspondingly, the configuration of 8×16×16 using eight ADiP cores of 16×16 achieves higher index than 2×64×64 and 4×32×32 configurations. A Legion with eight 16×16 ADiP cores is the optimal configuration, taking into consideration that workloads with a head size of 64 is efficiently processed and each Legion can instantiate 2× higher number of PEs than the 16×8×8 configuration.

\subsection{Block Matrix Multiplication with Parallel Cores}

Transformers and LLMs incorporate large-scale matrix multiplication workloads which are not feasible to be processed all at once. 
Block matrix multiplication is foundational in high-performance architectures, and basically partitions input matrices into smaller submatrices (blocks) \cite{golub2013matrix}. The multiplication of the entire matrices is reformulated as a series of smaller block-wise multiplications followed by summations of intermediate results. 
Modifying loop order and tile distribution substantially reshapes inter-core communication patterns and explores data reuse opportunities. 
The proposed accelerator adopts a column-major tile order to better align computation with column-wise data layout.

Each Legion orchestrates block-wise accumulation of the output product of input matrices in a column-wise manner with loop order $N \rightarrow K \rightarrow M$. 
Each K-chunk is decomposed across $C$ parallel cores, where each core receives tiles of input matrices and generates distinct psum tiles. Psum tiles are simultaneously and spatially reduced to a single tile, reducing psum memory access. 
This loop ordering enhances locality for the stationary matrix tiles and yields a compute-memory schedule well-suited for efficient matrix multiplication acceleration.

\section{D-Legion Architecture}

\begin{figure*}[t] 
  \centering
  \includegraphics[width=\textwidth, keepaspectratio]{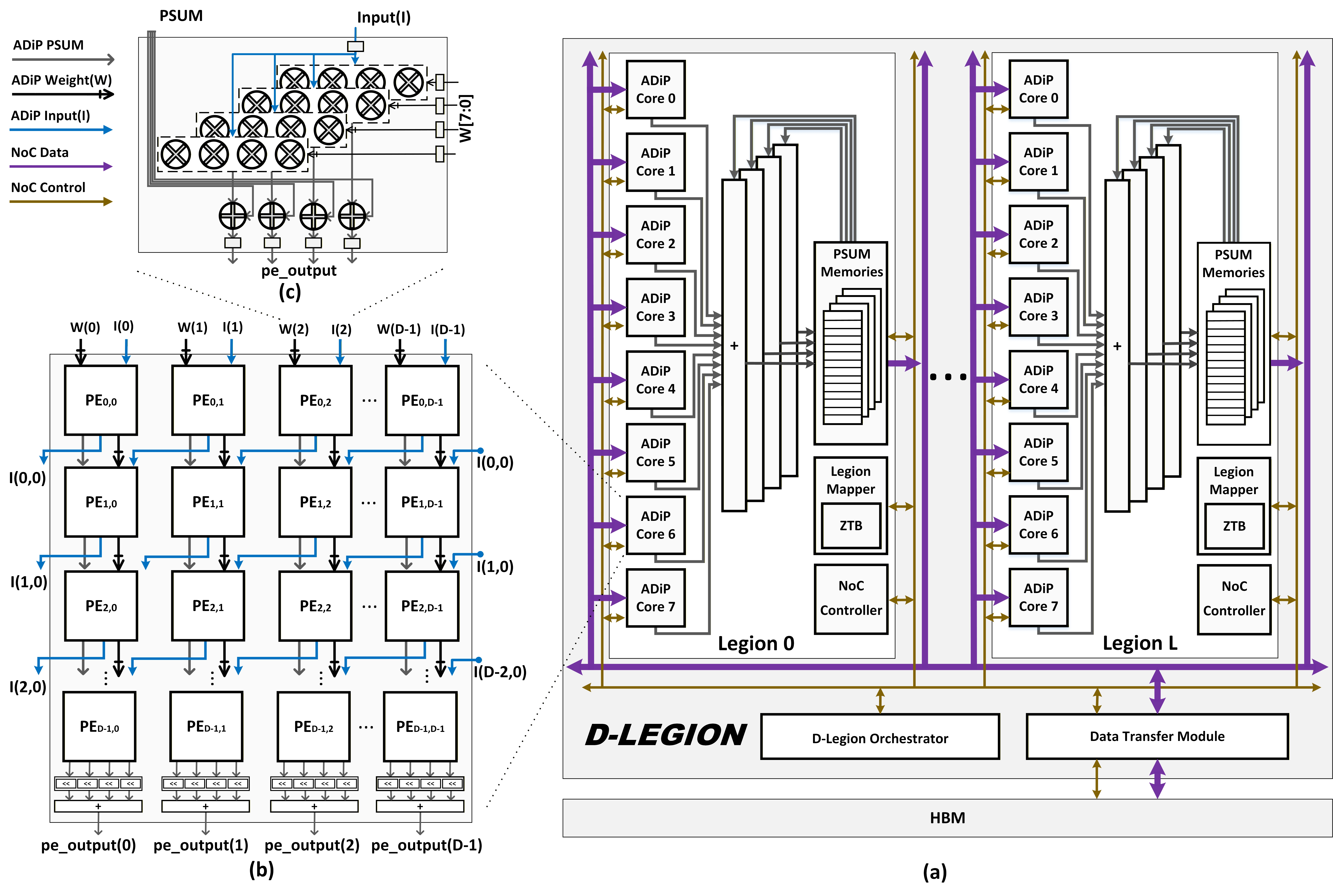} 
  \caption{(a) D-Legion architecture block diagram, consisting of $L$ Legions where each Legion has eight ADiP cores, accumulators, psum memories, Legion mapper, and local crossbar. (b) ADiP core block diagram, consisting of $D$×$D$ adaptive-precision systolic array. (c) Reconfigurable PE block diagram, consisting of 16 multipliers and arranged into four groups with their internal accumulators.}
  \label{Fig.5}
\end{figure*}

\begin{table*}[tb]
\centering
\caption{D-Legion Architecture Features and Specifications}
\label{tab:d_legion_features}
\begin{tabular}{p{0.1\textwidth}p{0.08\textwidth}p{0.7\textwidth}}
\toprule
\textbf{Feature} & \multicolumn{2}{l}{\textbf{Description}} \\
\midrule
\multirow{5}{=}{\textbf{Architecture \\ Organization}} 
& \multicolumn{2}{p{0.69\linewidth}}{D-Legion consists of $L$ Legions ($L=8$)} \\
& \multicolumn{2}{p{0.69\linewidth}}{Each Legion have $C$ ADiP cores ($C=8$)} \\
& \multicolumn{2}{p{0.69\linewidth}}{Each ADiP core is ($D\times D$) adaptive-precision systolic array ($D=16$).} \\
& \multicolumn{2}{p{0.69\linewidth}}{Each reconfigurable PE comprises 16 2-bit multipliers.} \\
& \multicolumn{2}{p{0.76\linewidth}}{NoC Hierarchy: Flexible NoC in addition to local crossbar (Cores$\leftrightarrow$Accumulators, Accumulators$\leftrightarrow$Psum memories).} \\
& \multicolumn{2}{p{0.69\linewidth}}{Shared links with Link\_ID to select between (weights, activations, and psums).} \\
\midrule
\multirow{2}{=}{\textbf{Adaptive-Precision PEs}} 
& \multicolumn{2}{p{0.69\linewidth}}{Activation-to-activation workloads are performed in INT8 precision.} \\
& \multicolumn{2}{p{0.8\linewidth}}{Projection workloads in 1-bit LLMs, where weights are represented in 2-bit, are accelerated by \textbf{4×} using ADiP cores.} \\
\midrule
\multirow{6}{=}{\textbf{Partial \\ Summations \\ Reduction}} 
& Spatial & 1$^{st}$ Level: Psums are reduced using ADiP cores. \\
& reduction & 2$^{nd}$ Level: Psums are reduced out from Legion's cores using parallel element-wise accumulators. \\
\cmidrule(lr){2-3}
& Temporal & 1$^{st}$ Level: Psums are reduced using ADiP cores. \\
& reduction & 2$^{nd}$ Level: Read the saved psum tile, accumulate the new psum tile, then save it to memory (row-wise). \\
\midrule
\multirow{6}{=}{\textbf{Maximized \\ Data Reuse}} 
& Inter-legion & For projection workloads: ADiP cores provide \textbf{4×} better data reuse. \\
\cmidrule(lr){2-3}
& \multirow{4}{=}{Intra-Legion} & The input matrix tiles are broadcasted to all Legions during the computations. \\
& & GQA and MQA attention layers with KV Reuse: for activation-to-activation workloads, KV matrix tiles are broadcasted to all corresponding Legions.   \\
\midrule
\multirow{4}{=}{\textbf{Zero \\ Multiplication \\ Detection}} 
& Scope & Block structured sparsity \\
\cmidrule(lr){2-3}
& Per core & The core is deactivated if the stationary matrix tile is zero to increase energy savings. \\
\cmidrule(lr){2-3}
& Per window & ($C$ tiles/Legion): The computations for this window are skipped, leading to higher acceleration, memory efficiency, and energy savings \\
\bottomrule
\end{tabular}
\end{table*}

This section presents the proposed D-Legion architecture. 
The D-Legion architecture exemplifies a computational hierarchy designed to exploit diverse optimization opportunities inherent in Transformers and LLMs. The proposed architecture comprises \textit{L} Legions, each instantiating eight ADiP cores configured as 16×16 systolic arrays, as shown in Fig. \ref{Fig.5}(a). The architectural design incorporates sophisticated strategies to maximize data reuse, and accelerate attention workloads in quantized Transformers and LLMs. 
Table \ref{tab:d_legion_features} summarizes the architecture organization and features.
The proposed architecture enables adaptive-precision, wherein the activation-to-activation mode leverages INT8 operations, the projection (activation-to-weight) mode exploits quantized multiplication, yielding up to quadrupled throughput. 
Psum reduction employs a two-tier reduction hierarchy encompassing both spatial and temporal strategies. The PEs of ADiP cores spatially and temporally reduce psums. Then, a second-level spatial reduction involves the Legion accumulators to reduce $C$ psum tiles to one psum tile. After that temporal reduction requires an incremental accumulation, involving iterative tile read-modify-write (RMW) sequences, after initially accumulating them in the cores.
For inter-Legion data reuse, each core increases data reuse during the processing of projection workloads. Intra-Legion strategies include sharing the input matrix with all heads split legion-wise during QKV projection or activation-to-activation workloads, leading to 8× higher data reuse compared to a single Legion. Additionally, they exploit GQA and MQA attention layers by replicating KV tiles across all Legions, achieving reuse factors proportional to the ratio of attention heads to KV heads. 
Moreover, output projection workload is split legion-wise, delivering 8× higher data reuse. 
Furthermore, the architecture explores block structural sparsity through zero-multiplication detection, enabling selective core deactivation and computational window bypassing when encountering zero tiles, thereby reducing energy consumption and accelerating the processing, simultaneously. 
Collectively, adaptive-precision, data reuse techniques, and sparsity-aware execution establish D-Legion as a domain-specialized accelerator optimized and tailored for attention workloads in quantized LLMs.

\subsection{Legion Micro-architecture}

Each Legion accelerates a separate matrix multiplication workload. Each Legion consists of eight cores, accumulators, psum memories, and a Legion mapper, as shown in Fig. \ref{Fig.5}(a). Each Legion acts as a dedicated accelerator unit and operates independently. All Legions receive only high-level workload dimensions from the global D-Legion orchestrator, allowing multiple Legions to progress concurrently across large-scale attention workloads with minimal cross-dependence. This hierarchy enables linear-scaling of the number of Legions with predictable performance and memory access patterns.

\subsubsection{ADiP Cores}

Each ADiP core is a $D \times D$ adaptive-precision systolic array incorporating reconfigurable PEs, shared shifters, and shared accumulators, as shown in Fig. \ref{Fig.5}(b) \cite{abdelmaksoud2025adip}. 
By adopting the DiP dataflow, ADiP cores adapt to different precisions and eliminate input and output synchronization FIFOs required by WS architecture \cite{abdelmaksoud2025dip}.
Each PE consists of 16 2-bit multipliers arranged into four groups with their internal accumulators, as shown in Fig. \ref{Fig.5}(c). The configuration of 16 multipliers is selected to process 8b×8b multiplications with a latency of one cycle.
These cores support multiple computation modes: projection mode (8b×4b or 8b×2b interleaved multiplications across two or four distinct tiles) and dense mode (8b×8b multiplication). In projection mode, single activation tile is shared across interleaved weight tiles. 
This configuration yields up to 4× higher computational throughput and correspondingly reduced input matrix fetching.

\subsubsection{Legion Accumulators}

Each Legion contains four accumulators, each implementing parallel element-wise adders that reduce psum tiles from eight spatial cores. During projection mode, a stream of up to 4×$D$ interleaved output elements from each core is forwarded to the parallel accumulators to be spatially reduced, before saving them to psum memories. However, during activation-to-activation mode, $D$ output elements (each is 32-bit) from each core are forwarded to the parallel accumulators. Two accumulators are tied together to add 32-bit inputs. Psums are accumulated row-wise. The accumulator controller is responsible for the operation mode, reduction sequencing, and initiation of psum accumulation. 

\subsubsection{Psum Memories}

Each Legion has four psum memory banks acting as a psum scratchpad used for accumulating intermediate psums. During projection mode, all psum banks are active for psum RMW operations. However, for activation-to-activation mode, only one bank is active, and the bank controller iterates over psum banks. Psum banks are carefully dimensioned so that they remain fully-local to each Legion during processing, avoiding cross-Legion memory interference. Each bank has a size of 0.66 MB with a total on-chip memory size of 2.64 MB.

\subsubsection{Legion Mapper}

Each Legion has a mapper, responsible for orchestrating tile progression through matrix dimensions ($M$,$K$,$N$). It performs sparsity-aware scheduling using a per-Legion zero-tile book (ZTB), a bitmask table that records which windows of tiles are structurally zero, as determined offline. Each window corresponds to $C$ tiles (one per core). A single bit per tile indicates whether it is structurally zero. 
The mapper uses these zero masks to cancel weights and activation transfers, disable cores, and skip accumulator updates when all tiles associated with a window are zero (fully-sparse windows). Alternatively, for partially-sparse windows, it deactivates the corresponding cores. Consequently, D-Legion avoids unnecessary computation and memory access.
The Legion controller is responsible for mapping the workload and controlling the computation flow. Additionally, it is responsible for zero detection, mode selection (activation-to-weight or activation-to-activation), core control, and psum memory control.

\subsection{D-Legion network-on-chip (NoC)}

The NoC provides flexibility for unicast/multicast matrix tiles to the target Legions/cores. It supports deterministic tile-level routing based on address prefixes, \texttt{[LEGION\_ID | CORE\_ID | LINK\_ID]}. The NoC enables selective multicast to multiple Legions, allowing KV tile sharing in GQA/MQA attention layers or workload partitioning to reduce off-chip memory access. For attention workloads, the NoC significantly reduces the memory footprint of attention computations by allowing KV tiles to be fetched once and replicated across relevant Legions eliminating redundant off-chip accesses.
Each Legion has a local crossbar connecting cores to accumulators and accumulators to psum memories. Additionally, all cores have shared Links with link IDs to select between (weights, activations, and psums). A NoC gateway handles tile payloads from the NoC.

\subsection{D-Legion Orchestrator and Workloads Mapping}

Each Legion executes a separate workload represented in ($M$,$K$,$N$) assigned by D-Legion orchestrator, operating independently of other Legions in terms of tile scheduling, accumulation, and memory. The D-Legion orchestrator is responsible for managing the processing of attention layers. 
The mapping enables the acceleration of attention layers as follows:

\begin{itemize}
    \item Mapping MHA workloads: Q,K,V projection workloads are mapped by assigning each head workload to one Legion. For activation-to-activation workloads, each workload is partitioned over all Legions along matrix dimension $N$, and D-Legion iterates over all heads. 
    \item Mapping GQA/MQA workloads: Similar to MHA workloads; however, activation-to-activation workloads: A group of Legions share the tiles of KV matrices.
    \item Workload partitioning: Output projection workload does not involve multi-head processing. Instead, the workload is partitioned across all Legions along matrix dimension $N$.
\end{itemize}

\section{Evaluation \& Results}

This section presents the evaluation of the D-Legion architecture across 1-bit LLM attention workloads in terms of latency, throughput, memory access, psum memory access.

\begin{figure}[t]
  \centering
  \includegraphics[width=\linewidth, keepaspectratio]{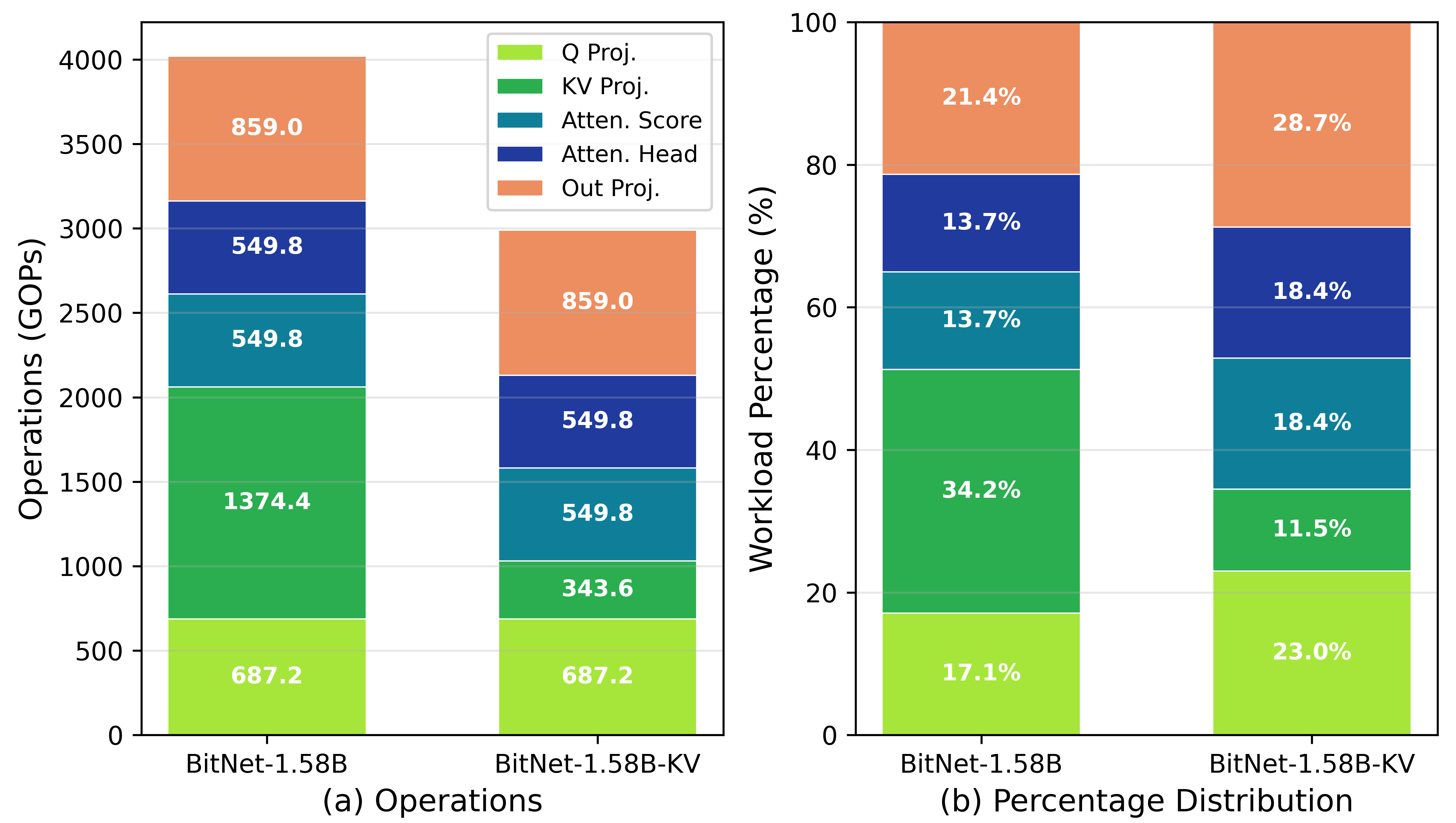} 
  \caption{Attention workloads distribution for the evaluated models: BitNet-1.58B and BitNet-1.58B-KV.}
  \label{Fig.6}
\end{figure}

\begin{figure*}[t]
  \centering
  \includegraphics[width=\linewidth, keepaspectratio]{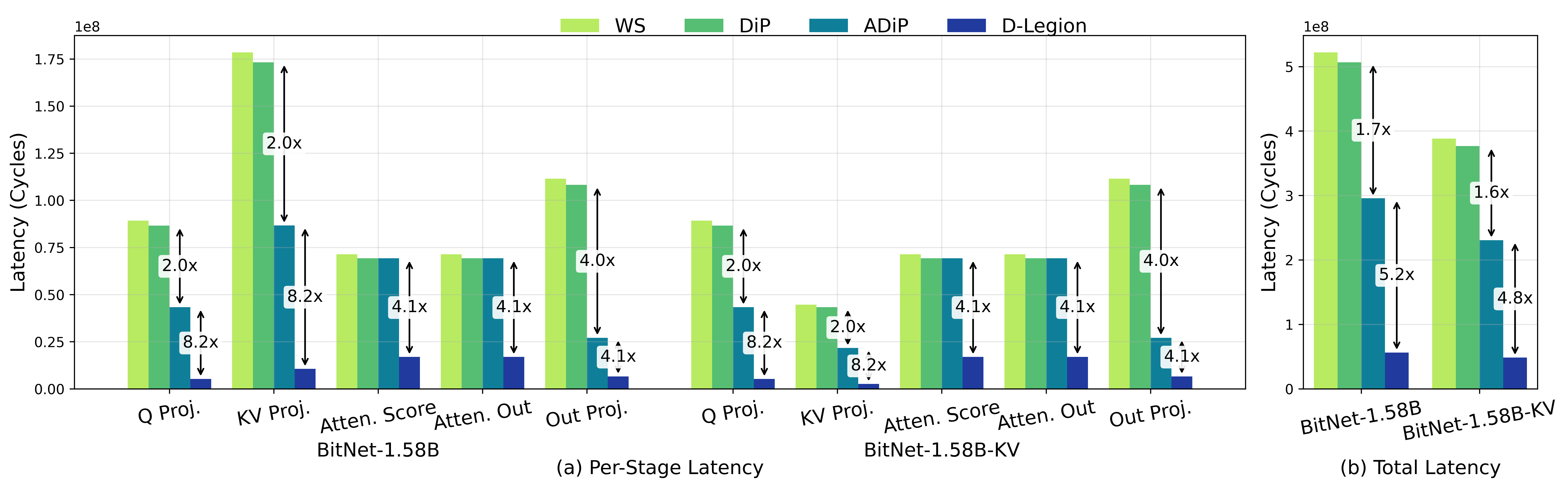} 
  \caption{Latency comparison across four hardware architectures (WS, DiP, ADiP, and D-Legion) for BitNet-1.58B and BitNet-1.58B-KV models. (a) Per-stage latency breakdown for attention layer stages. (b) Total latency per model for the attention workloads, highlighting the latency improvements achieved by D-Legion.}
  \label{Fig.7}
\end{figure*}

D-Legion is evaluated on attention workloads from two 1-bit LLM models: BitNet-1.58B with MHA layers and BitNet-1.58B-KV with GQA layers \cite{diao2023bitnet, wang20241}. 
BitNet-1.58B is a decoder model configured with 32 layers, a hidden size of 2560, 16 attention heads (each with a dimension of 128) and a maximum sequence length of 2048. It employs 2-bit weights, resulting in a compact architecture with an attention workload size (multiplications and additions) of nearly 4.02 TOPS.
BitNet-1.58B-KV is similar to BitNet-1.58B; however, the attention layers are GQA instead of MHA layers with 4 KV heads. The attention workload size is nearly 2.99 TOPS. The workload distribution and size for each stage are shown in Fig. \ref{Fig.6}. 
In D-Legion, attention score and attention output (activation-to-activation) workloads are performed in 8b×8b precision, while projection (activation-to-weight) workloads are performed in 8b×2b precision.

\subsection{Comparison with WS, DiP, and ADiP architectures}

The evaluation compares D-Legion with three hardware architectures: WS, DiP, and ADiP with a single-core of 64×64 with 4,096 PEs. The D-Legion architecture instantiates 16,384 PEs arranged in eight Legions, each with eight cores. The architecture organization enables scalable linear-scale performance with all cores being fully-utilized. A cycle-accurate simulator is developed to calculate the latency, throughput, and memory access for all architectures. 

Fig. \ref{Fig.7} presents the latency comparison of the proposed D-Legion architecture with three state-of-the-art architectures (WS, DiP, and ADiP) across two BitNet model variants: BitNet-1.58B and BitNet-1.58B-KV. 
Fig. \ref{Fig.7}(a) shows the latency breakdown per attention stage (cycles), revealing the architectural improvements. 
D-Legion achieves and up to 16.87× and 16.4× lower latency for QKV and output projections workloads compared to WS and DiP architectures, respectively.
Furthermore, D-Legion achieves up to 8.2× lower latency compared to ADiP architecture. 
WS and DiP consistently incur the highest latencies across all stages, as they perform the operations in INT8 precision, even for the quantized projection workloads. 
For ADiP with a single core of size 64×64, QKV projection workloads are limited by the head dimension of 128. 
Since the core size is 64×64, the latency is reduced by 2× instead of 4×.
D-Legion shows the lowest latency across all stages, with substantial improvements in both projection and activation-to-activation workloads, highlighting the architectural benefits. Fig. \ref{Fig.7}(b) shows the total latency, reinforcing the stage-level results: the latency for BitNet-1.58B-KV is reduced compared to BitNet-1.58B across all architectures, as KV heads are 4× lower than attention heads. D-Legion achieves up to 9.26×, 8.84×, and 5.2× lower latency compared to WS, DiP, and ADiP architectures, respectively.

\begin{figure*}[tb]
  \centering
  \includegraphics[width=\linewidth, keepaspectratio]{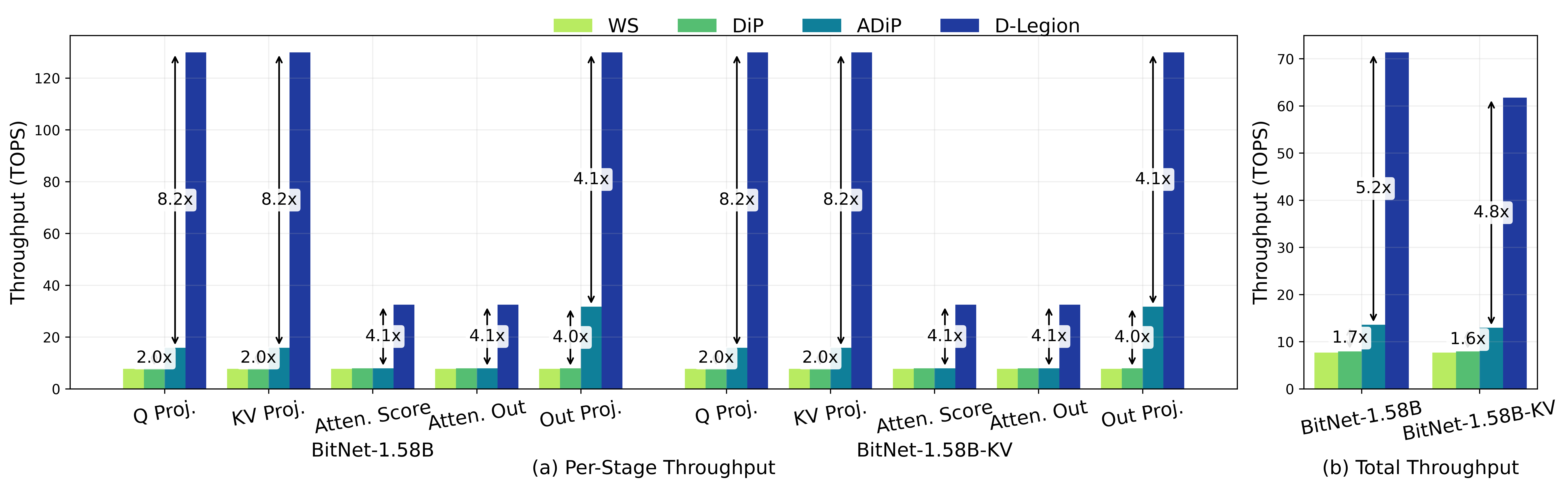} 
  \caption{Throughput comparison across four hardware architectures (WS, DiP, ADiP, and D-Legion) for BitNet-1.58B and BitNet-1.58B-KV models at a frequency of 1 GHz. (a) Per-stage throughput breakdown for attention layer stages. (b) Throughput per-model breakdown for the attention workloads, highlighting the throughput improvements achieved by D-Legion.}
  \label{Fig.8}
\end{figure*}


Figure \ref{Fig.8} presents the comparative analysis of the throughput for the evaluated architectures (WS, DiP, ADiP, and D-Legion) across attention workloads of two BitNet model variants: BitNet-1.58B and BitNet-1.58B-KV. 
Fig. \ref{Fig.8}(a) shows the throughput breakdown per attention stage. Relatively to the latency breakdown, D-Legion shows the highest throughput across all stages because the core size are selected to get the maximum computational throughput for the quantized workloads, and the PEs are organized in Legion fashion to provide the highest core utilization and process multiple parallel workloads, simultaneously. 
As a result, this enables instantiating 4× higher number of PEs without under-utilizing the cores. 
D-Legion runs at a frequency of 1 GHz. The projection workloads achieve up to 135.68 TOPs, and activation-to-activation workloads achieve up to 33.92 TOPs.
On the other hand, WS and DiP have the lowest throughput across all stages, as they perform operations in INT8 precision, even for the quantized projection workloads. 
Relatively to the core size, the throughput for ADiP is increased by only 2× compared to WS and DiP architectures.  
Fig. \ref{Fig.8}(b) shows the throughput analysis for the attention workloads per model. 
D-Legion achieves up to 9.26×, 8.84×, and 5.2× higher throughput for the attention workloads of BitNet-1.58B and BitNet-1.58B-KV models compared to WS, DiP, and ADiP architectures, respectively. 
The results show that D-Legion delivers exceptional throughput compared to WS, DiP, and ADiP architectures.

\begin{figure*}[tb]
  \centering
  \includegraphics[width=\linewidth, keepaspectratio]{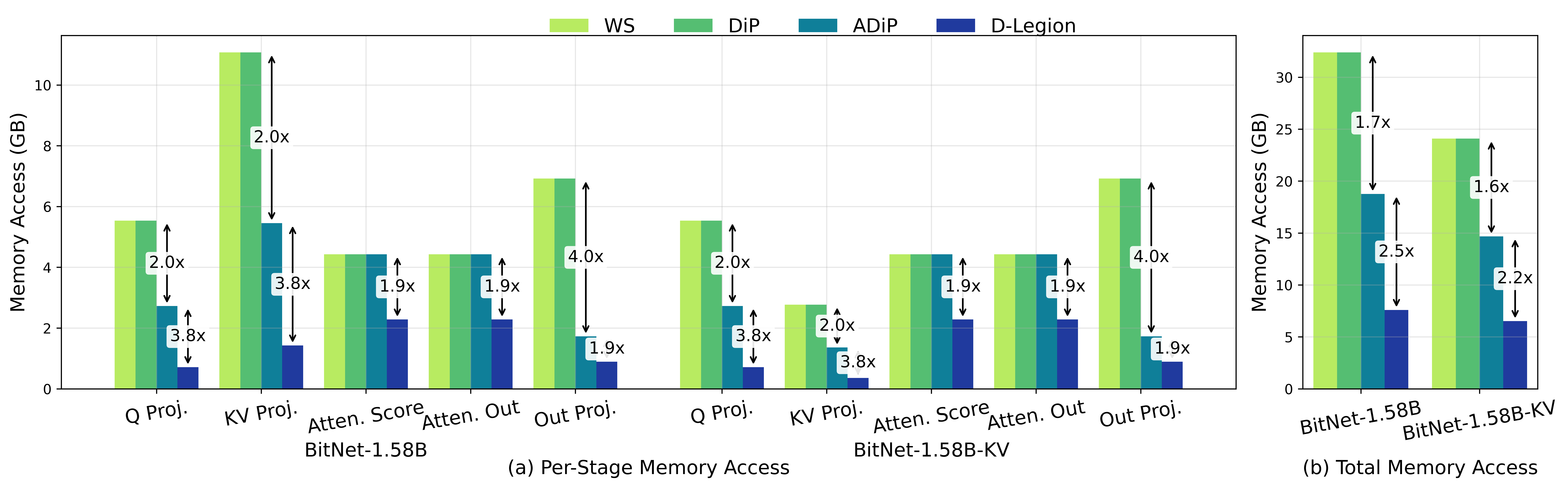} 
  \caption{Memory access comparison across four hardware architectures (WS, DiP, ADiP, and D-Legion) for BitNet-1.58B and BitNet-1.58B-KV models. (a) Per-stage memory access breakdown for attention layer stages. (b) Total memory access per model for the attention workloads, highlighting memory access savings achieved by D-Legion.}
  \label{Fig.9}
\end{figure*}

Figure \ref{Fig.9} compares memory access (weights and activations) across the evaluated architectures (WS, DiP, ADiP, and D-Legion) for the attention workloads of BitNet-1.58B and BitNet-1.58B-KV models. 
Fig. \ref{Fig.9}(a) shows the breakdown of memory access per-stage for the stages of the attention layers in gigabytes (GB). 
The analysis includes memory accesses associated with weights and activations loading.
D-Legion consistently offers the lowest memory access across all stages due to the architectural benefits achieved by the flexible NoC design and adopted data reuse strategies. 
The projection workloads exhibit the least memory access due to inter-Legion and intra-Legion data reuse up to 3.8× compared to ADiP architecture, and up to 7.6× compared to WS and DiP architectures. 
Additionally, memory access is reduced for activation-to-activation workloads due to intra-Legion data reuse with up to 1.9× compared to other architectures.  
Fig. \ref{Fig.9}(b) shows the total memory access for the attention workloads per model, showing up to 2.5× and 4.25× lower memory access for D-Legion compared to ADiP and DiP architectures, respectively.

\begin{figure*}[tb]
  \centering
  \includegraphics[width=\linewidth, keepaspectratio]{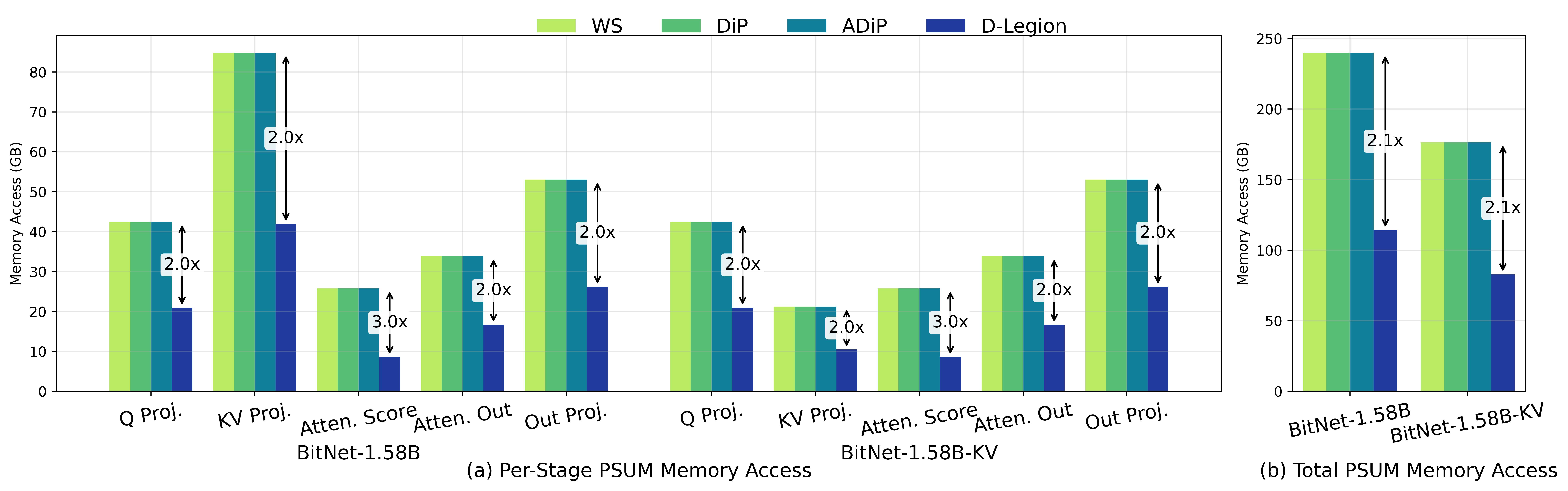} 
  \caption{A comparison of psum memory access across four hardware architectures (WS, DiP, ADiP, and D-Legion) for BitNet-1.58B and BitNet-1.58B-KV models. (a) Per-stage psum memory access breakdown for attention layer stages. (b) Total psum memory access per model for the attention workloads, highlighting psum memory access savings achieved by D-Legion.}
  \label{Fig.10}
\end{figure*}

Finally, the results presented in Fig. \ref{Fig.10} illustrate the comparative analysis of psum memory access across the evaluated architectures (WS, DiP, ADiP, and D-Legion). In D-Legion, each Legion incorporates four parallel accumulators to spatially reduce psums. 
Fig. \ref{Fig.10}(a) presents the breakdown of psum memory access for the stages of the attention layers, showing up to 3× lower psum memory access for D-Legion compared to WS, DiP, and ADiP architectures.
As a result, total psum memory access is reduced by up to 2.1× for D-Legion compared to other architectures, as shown in Fig \ref{Fig.10}(b). 
Thus, energy consumption for reading and writing psums is reduced.

\subsection{D-Legion Architecture Scaling}

D-Legion is a scalable architecture that offers linear-scale performance with the number of Legions. As discussed earlier, each Legion accelerates a separate workload per head or partitions large-scale workloads across multiple Legions. The baseline D-Legion architecture consists of eight Legions; however, the number of Legions can be increased to achieve higher acceleration with respect to the memory specifications. 

High bandwidth memory (HBM) has advanced through several generations providing higher bandwidth, and increased memory size. HBM3 raises aggregate bandwidth up to 819 GB/s per stack and supports up to 16 stacks, with total capacity reaching up to 64 GB \cite{jedec_hbm3_std}. The expanded memory storage and increased memory bandwidth mitigate the compute-intensive and memory-bound challenges of modern AI models. 

In the D-Legion architecture, each Legion requires a 1024-bit interface operating at 1 GHz, providing a bandwidth of 128 GB/s. The architecture can be scaled up to 64 Legions using 16 HBM3 stacks, each providing 512 GB/s of bandwidth.
As a result, D-Legion with 64 Legions achieves a peak throughput of 1085.44 TOPS.

\subsection{Comparison with Google TPUv4i}

\begin{figure*}[tb]
  \centering
  \includegraphics[width=\linewidth, keepaspectratio]{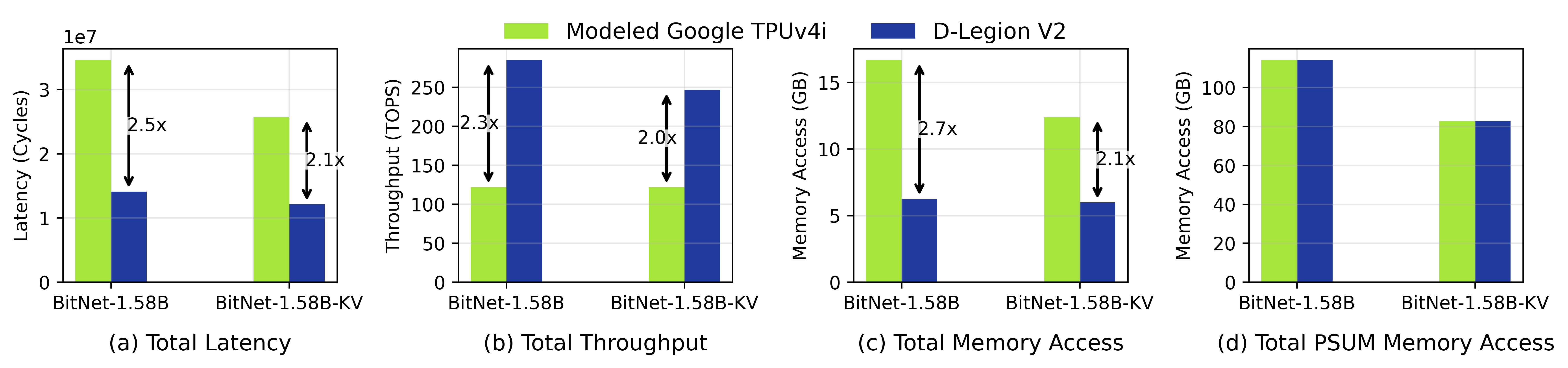} 
  \caption{A comparison between modeled Google TPUv4i and D-Legion V2 with the same number of PEs using attention workloads from BitNet-1.58B and BitNet-1.58B-KV models.
  (a) An analysis of total latency per model. (b) An analysis of total throughput per model at a frequency of 1 GHz. (c) An analysis of total memory access (GB) per model. (d) Total psum memory access (GB) per model.}
  \label{Fig.11}
\end{figure*}

Google TPUv4i consists of 4 matrix multiply units (MXUs), each is 128×128 systolic array with total 65,536 PEs, running at a frequency of 1.05 GHz \cite{jouppi2021tpuv4i}. 
D-Legion is compared to modeled Google TPUv4i to evaluate the latency, throughput, memory access, and psum memory access, as shown in Fig. \ref{Fig.11}. 
As the baseline D-Legion architecture consists of eight Legions with total number of PEs equals 16,384, the number of Legions is increased to 32 Legions to instantiate the same number of PEs as Google TPUv4i. 
D-Legion V2 achieves up to 2.5× lower total latency compared to Google TPUv4i, as shown in Fig. \ref{Fig.11}(a).
Furthermore, D-Legion V2 achieves up to 2.3× higher throughput compared to Google TPUv4i, as shown in Fig. \ref{Fig.11}(b).
As shown in Fig. \ref{Fig.11}(c), D-Legion V2 reduces total memory access by up to 2.7× compared to Google TPUv4i.
Finally, total psum memory access is the same for both architectures, as shown in Fig. \ref{Fig.11}(d).

\section{Conclusion}

This paper presents D-Legion, a novel scalable many-core adaptive architecture, designed using many adaptive-precision systolic arrays to accelerate matrix multiplication for quantized LLMs.
The proposed architecture consists of $L$ Legions where each Legion has $C$ cores of adaptive-precision systolic arrays. 
A comprehensive design space exploration is performed in terms of Legion/core granularity to determine the optimal architecture configuration.
The proposed architecture supports multiple computation modes, including quantized sparse and dense matrix multiplication workloads for different attention types such as MHA and GQA layers.
The proposed architecture exploits block structured sparsity within a fully-sparse or partially-sparse windows. 
In addition, psum memory access is spatially reduced using parallel element-wise accumulators per each Legion. 
Furthermore, data reuse is maximized through optimized scheduling techniques by multicasting matrix tiles with KV reuse, or workload partitioning. 
D-Legion is evaluated on attention workloads from two 1-bit LLM (BitNet) models, delivering up to 8.2× lower latency, 3.8× higher memory savings, and 3× higher psum memory savings compared to state-of-the-art work. 
D-Legion, with eight Legions and 64 total cores, achieves a peak throughput of 135.68 TOPS at a frequency of 1 GHz.
Furthermore, the scalability of D-Legion architecture is studied. Based on the memory specifications, the number of Legions per architecture can be increased up to 64 Legions, delivering up to 1085.44 TOPS.
Moreover, D-Legion with 32 Legions is compared to Google TPUv4i, achieving up to 2.5× lower total latency, up to 2.3× higher throughput, and up to 2.7× higher total memory savings for attention workloads from two BitNet models.
This paper is the foundation for D-Legion architecture. Future work will focus on physical hardware implementation, and development of fine-tuned LLM models employing block structured sparsity.

\bibliographystyle{IEEEtran}
\bibliography{references}

\begin{IEEEbiography}
[{\includegraphics[width=1in,height=1.2in,clip,keepaspectratio]{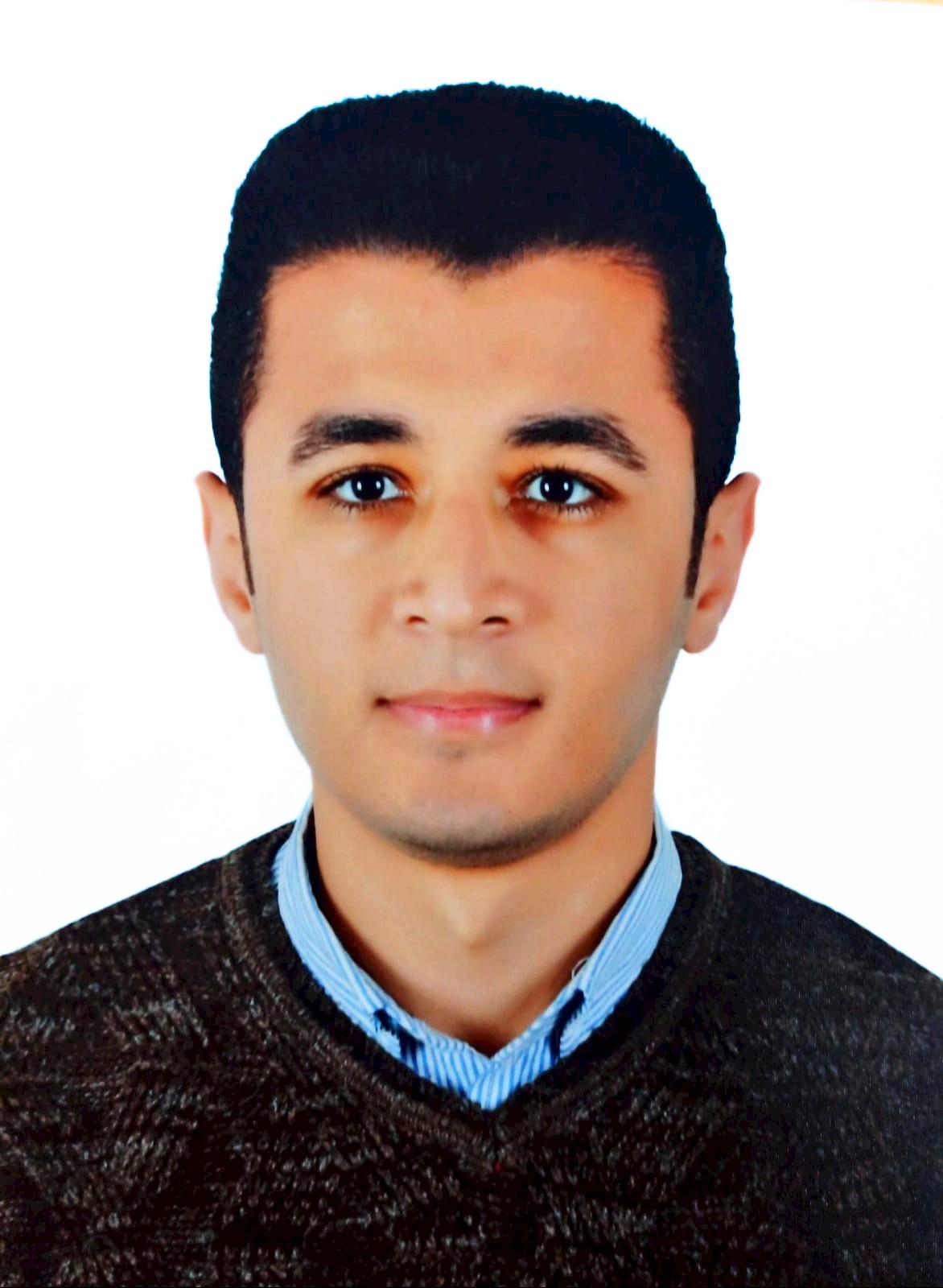}}]{Ahmed J. Abdelmaksoud}
 is currently pursuing his PhD with the Centre for Electronics Frontiers (CEF) at the University of Edinburgh, UK. He received his BSc and MSc in Electronics Engineering from Cairo University, Egypt in 2018 and 2022, receptively. Since 2018, he has been actively involved in Digital ASIC design projects across both research and industry. His professional experience includes working as a Research Associate at the System-on-Chip Center, Khalifa University, UAE; an ASIC Physical Design Engineer at Si-Vision, Egypt; and a Research Assistant at the Opto-Nano Electronics Lab, Egypt. In addition, his current research interests primarily focus on developing spatial and specialized architectures for efficient AI hardware acceleration.
\end{IEEEbiography}
\vspace{-10pt}
\begin{IEEEbiography}
[{\includegraphics[width=1in,height=1.2in,clip,keepaspectratio]{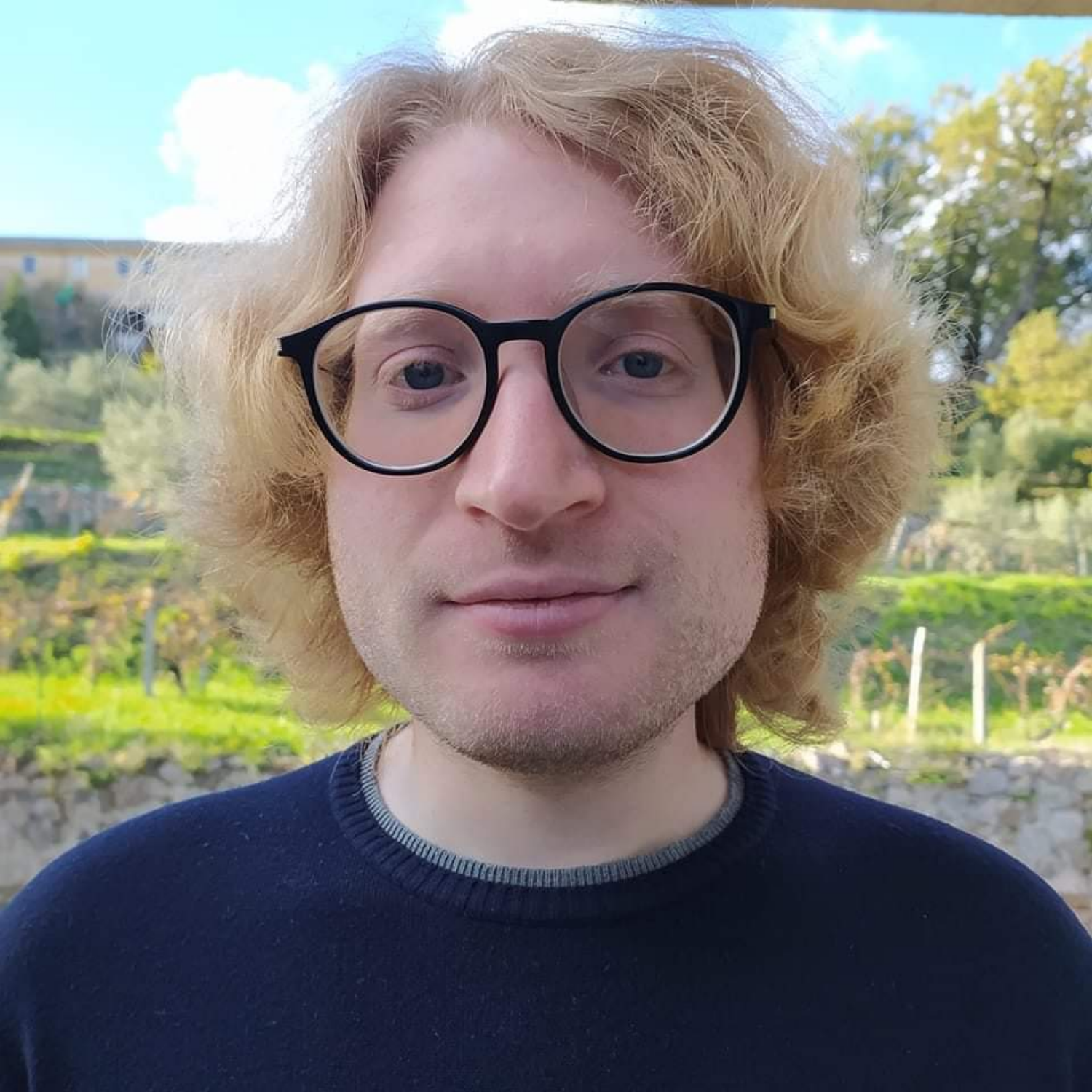}}]{Cristian Sestito}
(Member, IEEE) is a Research Fellow at the Centre for Electronics Frontiers CEF, The University of Edinburgh (UK). He received his BSc and MSc degree from University of Calabria (Italy), both in Electronic Engineering. He got his PhD in Information and Communication Technologies from the same university in 2023, focusing on Convolutional Neural Networks and their implementation on Field Programmable Gate Arrays (FPGA). In 2021/2022, Cristian was a Visiting Scholar at Heriot-Watt University, Edinburgh, working on neural networks’ compression. His research interests include digital design, embedded system design for AI on FPGA-based systems-on-chip, software simulators for neuromorphic AI, AI for circuits and systems design automation.
\end{IEEEbiography}
\vspace{-10pt}
\begin{IEEEbiography}
[{\includegraphics[width=1in,height=1.2in,clip,keepaspectratio]{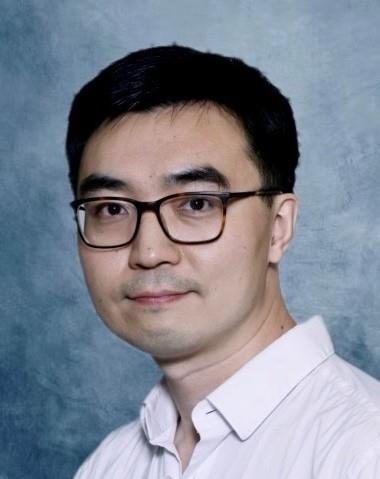}}]{Shiwei Wang} 
(Senior Member, IEEE) received the B.Eng. degree (Hons.) in electronic engineering from Zhejiang University, China, in 2010, and the Ph.D. degree in microelectronics from The University of Edinburgh, U.K., in 2014. He was a Research Assistant Professor at Shenzhen Institute of Advanced Technology, Chinese Academy of Sciences, China, from 2014 to 2015, a Senior Researcher at imec, Belgium, from 2015 to 2020, and an Associate Professor at the Department of Electronics and Computer Science, University of Southampton, U.K., from 2020 to 2022. In 2022, he joined the School of Engineering, The University of Edinburgh, where he is currently a Reader and a Bayes Innovation Fellow. His research interests include analog and mixed-signal integrated circuits for emerging application including implantable/wearable electronics, brain–machine interface, and sensor instrumentation.
\end{IEEEbiography}
\vspace{-10pt}
\begin{IEEEbiography}
[{\includegraphics[width=1in,height=1.2in,clip,keepaspectratio]{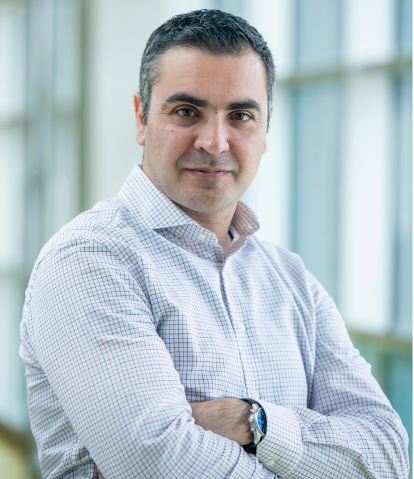}}]{Themis Prodromakis}
(Senior Member, IEEE) received the bachelor’s degree in electrical and electronic engineering from the University of Lincoln, U.K., the M.Sc. degree in microelectronics and telecommunications from the University of Liverpool, U.K., and the Ph.D. degree in electrical and electronic engineering from Imperial College London, U.K. He then held a Corrigan Fellowship in nanoscale technology and science with the Centre for Bio-Inspired Technology, Imperial College London, and a Lindemann Trust Visiting Fellowship with the Department of Electrical Engineering and Computer Sciences, University of California at Berkeley, USA. He was a Professor of nanotechnology at the University of Southampton, U.K. He holds the Regius Chair of Engineering at the University of Edinburgh and is Director of the Centre for Electronics Frontiers. He is currently a Royal Academy of Engineering Chair in emerging technologies and a Royal Society Industry Fellowship. His background is in electron devices and nanofabrication techniques. His current research interests include memristive technologies for advanced computing architectures and biomedical applications. He is a fellow of the Royal Society of Chemistry, the British Computer Society, the IET, and the Institute of Physics.
\end{IEEEbiography}

\end{document}